\documentclass[12pt]{article}
\usepackage[dvips]{color}
\usepackage{amsmath}
\usepackage{graphicx}
\usepackage{subfigure}
\usepackage{epsfig}
\usepackage{amsmath}
\usepackage{cite}
\usepackage{color}
\usepackage{subfigure}

\begin{document}
\begin{center}
\Large{\bf Einstein-aether scalar-tensor anisotropic constant-roll inflationary scenario in
noncommutative phase space}\\
\vspace{.8cm}\small {\bf A. Pasqua$^{\dag}$\footnote {Email:~~~toto.pasqua@gmail.com}}, \hspace{.5cm}{\bf S. Noori Gashti$^{\star}$\footnote {Email:~~~saeed.noorigashti@stu.umz.ac.ir}}, \quad
\\
\vspace{0.5cm}$^{\dag}${Department of Physics, University of Trieste, Via Valerio, 2 34127 Trieste, Italy.}\\
\vspace{0.5cm}$^{\star}${Department of Physics, University of Mazandaran P. O. Box 47416-95447, Babolsar, Iran}\\
\small \vspace{1cm}
\end{center}
\begin{abstract}
The primary purpose of this study is to investigate the constant-roll inflationary scenario with anisotropic conditions concerning the Einstein-aether scalar-tensor cosmology in noncommutative phase space. We first introduce an Einstein-aether scalar-tensor cosmological model. In this structure, one can introduce an aether field with aether coefficients in the action integral of scalar-tensor. It will be a function of the scalar field, which is, in fact, a kind of extender of the Lorentz-violating theories. Hence, we present the point-like Lagrangian, which represents the field equations of the Einstein-aether scalar-tensor model. Then we calculate the Hamiltonian of our model directly. According to the noncommutative phase space characteristics, we will calculate the specific equations of this model. Then, according to the constant roll conditions, we take the anisotropic constant-roll inflationary scenario and calculate some cosmological parameters of the mentioned model, such as the Hubble parameter, potential, etc.\\\\
Keywords: Anisotropic constant-roll, Noncommutative parameter, Einstein-aether Scalar-tensor, inflationary scenario
\end{abstract}
\newpage
\tableofcontents

\section{Introduction}

One of the most famous problems researchers are trying to solve with different theories is quantum gravity. A common feature that can be found in all topics related to quantum gravity is called the Lorentz violation\cite{1}. However, various gravitational models that face the Lorentz violation have recently received a lot of attention\cite{2,3,4,5,6,7,8,9,10,11}.
One of the theories of gravity that has received much attention in connection with Lorentz violating is called the Einstein-aether theory \cite{12,13,14,15,16}.
there exist presented the quantities of the unitary time-like vector field, the aether field In the Einstein-Hilbert action Integral
Selecting this field leads to a violation of Lorentz symmetry in the preferred frame. The limitation of Einstein's General Relativity lives while the preservation of locality and covariance formulation is guaranteed in Einstein-aether theory.
The mentioned theory, Einstein-aether theory, is called a second-order theory, which is used to describe different gravitational systems\cite{17,18,19,20,21,22,23,24}.
Einstein-aether theory has many features, and the cosmological applications of this theory have been widely discussed in the literature, including the description of the classical limit of Horava-Lifshitz\cite{25}. Of course, the critical point here is that the opposite is not possible\cite{26,27}.
Of course, from another point of view, scalar fields play a vital role in describing the universe. The field used to describe the early acceleration era of the universe is known as the inflaton field.
In addition, scalar fields play a very important role in describing the late-time acceleration as the solution to the dark energy problem\cite{28,29,30,31,32,33,34,35}.
Another characteristic of the scalar field is that they attribute the degrees of freedom supplied by higher derivatives in gravitational action as a result of quantum corrections or modification derived from general relativity\cite{36,37,38}.
As a result, the various types of gravitational models that describe aether Lorentz violating inflationary solutions have been widely introduced by researchers in the literature\cite{16,39}.
Among the many works done in this field are many analyses and evaluations that show that the Einstein-aether scalar-tensor model can explain more than one inflationary period, just as, unlike scalar-tensor theory, strong cosmological history. For further study and analysis in this field and the different types of theories and solutions presented in this framework, which have answered many questions, you can see in\cite{39,40,41,42,43,44,45}.
 This paper aims to present a new challenge that has not been explored so far, namely the constant-roll inflationary scenario for the Einstein-aether scalar-tensor model in the noncommutative phase space and compare the results with other theories mentioned in the literature. One of the theories that have led to the most important challenges in cosmology so far is called Einstein's theory of general relativity, which has undergone generalizations and modifications\cite{46,47}.
Among these modifications, which have exciting features and results, are modifications of the renormalizability of quantum field theory, which somehow encounters a particular framework that we discuss in this article, called noncommutative space-time.
Noncommutative phase space has been discussed in various theories of cosmology, and its various cosmological applications about different theories and frameworks have been discussed. The results have been compared with the latest observable data. You can see for farther study in \cite{47,48,49,50,51,52,53,54,55,56,57}. In many calculations, the effects of noncommutative parameters have been studied on various types of cosmological theories such as power-law inflation, measurement of CMB, etc.\cite{58,59,60,61,62,63,64,65}. In fact, it is basically universally accepted nowadays that quantum gravity is holographic so that the highest amount of data permitted in a given area of space-time is proportional to the border region rather than the region's volume. This issue is against the conventional image of extensivity of information (entropy) but in agreement with the proposal of Bekenstein on the proportionality of black hole entropy to its event horizon area. It may The notion of holography can be tied to the same causal system and therefore, to the equivalence principle and local Lorentz invariance. To that end, we consider a modified gravity theory called Einstein-aether theory. Einstein-aether theory violates local Lorentz invariance and therefore annihilates the notion of a universal light cone. Yet, the low energy limit contains static and spherically symmetric solutions with universal horizons-spacelike hypersurfaces that are causal borders between an interior area and asymptotic spatial infinity. Results suggest that the scope of holography may be much broader than currently assumed. However, one must go a long way before these questions are convincingly settled.
Researchers have recently studied the effects of noncommutative parameters on the inflation scenario of constant rolling in the face of various structures such as fermion systems, modified Brans-Dicke cosmology, and other cosmological forms; results are compared with the latest observable data\cite{66}.
The constant-roll inflation scenario has also been of great interest to researchers recently.
In this theory, instead of using slow-rolling formalism for inflationary studies, they use a special condition called constant-roll condition, which challenges inflation scenarios and provides analytical and accurate answers for some cosmological parameters, such as the Hubble parameter, potential, scale factor, and so on.
This theory has been widely discussed in the literature. Such a condition has challenged various structures of effective theories such as low energy effective theory, f(R) gravity, and other modified gravitational theories. The results with the latest observational data and accepted and other theories in the literature have yielded exciting results.
Some of these works can be found in\cite{67,68,69,70,71,72,73,74}. All of the above motivated the article to be organized in the following form.\\
 Section 2 reviews the basic concepts and equations of Einstein-aether Scalar-tensor Cosmology.
Section 3 describes the noncommutative phase space; then, we obtain the Hamiltonian of the Einstein-aether Scalar-tensor model. According to the properties and indices related to the noncommutative phase space, we calculate the specific equations of the mentioned model in noncommutative phase space in section 4.
In section 5, considering the constant-roll condition, we examine the anisotropic inflation scenario and calculate the analytical and precise answers for some cosmological parameters such as the Hubble parameter, potential, etc. Finally, we describe the results in detail in section 6.
\section{Einstein-aether scalar-tensor cosmology}

In this section, we first explain the basic concepts and equations of Einstein-aether scalar-tensor cosmology. Then, by briefly introducing the noncommutative phase space, we will explain how the mentioned model behaves in the noncommutative space and present the required equations such as field equations.
Recently, different types of Einstein-aether cosmological models with the scalar fields have been introduced, and some work has been done in this field in the literature\cite{16}.
Among them, the potential of a scalar field for the quintessence field is assumed as a function of specific aether field variables, and its structures are challenged, which has been studied as a general and basic model. Kanno and Soda \cite{39} with the introduction of specific Lagrange, attracted the attention of the scientific community.
They introduced an integral action concerning the Einstein-aether coupling parameters, a scalar field function.
Such a study led to the fact that this cosmic model experiences two periods of inflation. When the scalar field is dominant, we will have a slow-roll era, and when the aether field contributes to the cosmological fluid, the Lorentz violating state will be established. With respect to \cite{39} the results were presented, including the dynamics of the chaotic inflationary model. The interpretations were used to introduce toy models to study structures such as the Lorentz violating DGP model with no ghosts\cite{45}.
The above model has been extensively studied in the literature by researchers, and its various cosmological applications have been investigated, among which you can see in\cite{45,75,76}.
The Lorentz violating study has also been studied to analyze cosmological histories and cosmological observations and found that Einstein-aether cosmology can be used to describe cosmological observations \cite{75,77}.
Meanwhile, the study of the dynamics models by aether field has been the subject of work of many researchers, and a lot of work has been done that for further research you can see\cite{78,79,80,81,82,83,84}.
Some researchers have also studied Einstein-aether scalar field cosmology using exact symmetry analysis \cite{85}.
Among the most important work done by researchers in recent years, \cite{86} that have quantized in Einstein-aether scalar field cosmology.
Using the descriptions detailed in\cite{39} It is discussed that the generalization of the gravitational model can be considered and a scalar field assumed in the Jordan framework, i.e., a scalar field that is coupled with the gravitational section.
The Einstein-aether scalar-tensor gravitational model can be considered an integral action in the following form.

\begin{equation}\label{1}
\mathcal{S}=\mathcal{S}_{ST}+\mathcal{S}_{aether}
\end{equation}

We will now introduce each component separately; hence the scalar-tensor theory integral action with respect to\cite{87} It is written in the following form.

\begin{equation}\label{2}
\mathcal{S}_{ST}=\int d^{4}x\sqrt{-g}\bigg(F(\phi)R+\frac{1}{2}g^{\mu\nu}\phi_{;\mu}\phi_{;\nu}+V(\phi)\bigg)
\end{equation}

In the above equation, the parameters $F(\phi)$ and $V(\phi(x^{\nu}))$ specify the coupling function of the scalar field and the scalar field potential, respectively. Now the second part of equation 1 means $\mathcal{S}_{aether}$ with respect to\cite{39} is written in the following form.

\begin{equation}\label{3}
\begin{split}
\mathcal{S}_{aether}=&-\int d^{4}x\sqrt{-g}\bigg(\beta_{1}(\phi)u^{\nu;\mu}u_{\nu;\mu}+\beta_{2}(\phi)(g^{\mu\nu}u_{\mu;\nu})^{2}+\beta_{3}(\phi)u^{\nu;\mu}u_{\mu;\nu}\\
&+\beta_{4}(\phi)u^{\mu}u^{\nu}u_{;\mu}u_{\nu}-\lambda(u^{\mu}u_{\nu}+1)\bigg)
\end{split}
\end{equation}

In the above equation, we have a series of parameters such as $\beta_{1}$, $\beta_{2}$, $\beta_{3}$, and $\beta_{4}$ coefficient functions which describe the coupling between the aether field and the scalar field, and ($\lambda$) Lagrange multiplier, which specifies the unitarian of the aether field $(u^{\mu}u_{\mu}+1=0)$. Also, the homogeneous and isotropic universe is described using the FLRW flat metric according to the cosmological structures, which is mentioned in the following form.

\begin{equation}\label{4}
\begin{split}
ds^{2}=-N^{2}(t)dt^{2}+a^{2}(t)\big(dx^{2}+dy^{2}+dz^{2}\big)
\end{split}
\end{equation}

Here, $N(t)$ and $a(t)$ specify the lapse function and scale factor, respectively, which can explain the three-dimensional radius of the Euclidean space.
Another point to note here is that for a comoving observe, the expansion rate of the parameter $\theta$ is specified in the form $\theta=3H^{2}$, where $H$ is the Hubble parameter, obtained by dividing the scale factor derivative upon the scale factor itself.
According to equation (4), Ricci scalar is defined as form $R=12H^{2}+\frac{6}{N}\dot{H}$. We also assume the symmetries of the background space for the scalar field and the aether field $\phi=\phi(t)$ to explain that it guarantees the unitarity condition for the aether field $u^{\mu}=\frac{1}{N}\delta^{\mu}_{t}$.
According to equation (1), integrating parts of the gravitational Action and the above explanations, we will have a formal equation.

\begin{equation}\label{5}
\begin{split}
\mathcal{S}=\int \frac{dt}{N}\bigg(6F(\phi)a\dot{a}^{2}+6F_{,\phi}a^{2}\dot{a}\dot{\phi}+\frac{1}{2}a^{3}\dot{\phi}^{2}-N^{2}a^{3}V(\phi)+3\bigg[\beta_{1}(\phi)+3\beta_{2}(\phi)+\beta_{3}(\phi)\bigg]a\dot{a}^{2}\bigg)
\end{split}
\end{equation}

According to the above explanations and equations, the Lagrangian point-like equation of the mentioned model, which can describe the field equations, is written in the following form.
The same equation that we will vigorously challenge in continuing our calculations.

\begin{equation}\label{6}
\begin{split}
\mathcal{L}\big(N, a, \dot{a}, \phi, \dot{\phi}\big)=\frac{1}{N}\bigg(6A(\phi)a\dot{a}^{2}+6B(\phi)a^{2}\dot{a}\dot{\phi}+\frac{1}{2}a^{3}\dot{\phi}^{2}\bigg)-Na^{3}V(\phi)
\end{split}
\end{equation}

In the above equation, we have some vital parameters, including $A(\phi)=F(\phi)+\frac{1}{2}(\beta_{1}(\phi)+3\beta_{2}(\phi)+\beta_{3}(\phi))$ and $B(\phi)=F(\phi)_{,\phi}$.

One of the characteristics of gravitational models can be the minisuperspace description.
In the next section, we will describe the basic concepts and equations of the noncommutative space and then integrate these concepts.

\section{Noncommutative phase space}

In this section, we provide a detailed description of the noncommutative phase space. Since Einstein's theory of gravitation is not suitable for explaining the structures of the universe at very high energies, researchers have proposed alternatives that result from the modification or expression of theories with new systems.
In this regard, various theories and formulations can be mentioned, including the structure Snyder\cite{46,47} Which describes a specific set of the NC spacetime coordinates.
This structure introduces a short-length cutoff called the noncommutative parameter, which can modify features such as renormalizability properties of relativistic quantum field theory. \cite{48,49,50,88}.
NC effects can be significant when dealing with scales where the effects of quantum gravity are substantial.
Since the problem of cosmological inflation at such energy scales has significant challenges, the use of such deformed phase space scenarios at these scales seems appropriate to study such a dynamic phase of the universe.
In general, such a structure of spacetime has recently been considered by researchers, and many of its cosmological applications have been studied and compared with the latest observable data as well as other works in the literature.
For example, clearly in \cite{89} NC spacetime affected on power-law inflation has been investigated and showed a specific function of running the spectra index.
Also, in \cite{90}, the effects of noncommutativity on cosmic microwave background have been studied, and sufficient studies and calculations have shown that noncommutativity may cause the spectrum of fluctuation to be non-Gaussian and anisotropic.
Among other studies that have been done in this field, one can challenge the effect of the noncommutativity phase space on the spectral index, power spectrum, and running spectral index of the curvature perturbations in different structures in the inflationary universe. You can see \cite{89,90,91,92,93,94}.
As stated in the literature, different structures of noncommutativity space have been used to study the early universe. In this article, we also use a specific framework to examine the structure and scenario of inflation.
Thus, introducing noncommutativity characteristics in classical or quantum cosmology can offer an attractive incentive to explore new concepts.
In the present work, we take advantage of this feature. The effects of noncommutativity phase space are investigated by using classical canonical noncommutativity features into Poisson brackets.
We obtain here a specific type of these achievements, which is a particular type of canonical noncommutativity using an appropriate deformation on the classical phase space variables.
One of the most important relations that we will benefit from in the calculations is the deformed Poisson bracket between the canonical conjugate moment, expressed in the following relation.

\begin{equation}\label{7}
\begin{split}
\{P_{a}, P_{\phi}\}=\theta\phi^{3}
\end{split}
\end{equation}

In the above equation, the noncommutativity phase space parameter, denoted by ($\theta$), is assumed constant.
This linear suggestion is fascinating. NC ingredient has also been used in studies related to the gravitational collapse of a scalar field, which has yielded exciting results\cite{20}.
In general, we believe that such a choice between dynamic noncommutativity between the momenta supplies a much more attractive relationship for the dynamic study of the universe's evolution.
We can also apply the following formulas in our calculations.

\begin{equation}\label{8}
\begin{split}
\{P_{a}, f(P_{a},P_{\phi})\}=\theta\phi^{3}\frac{\partial f}{\partial P_{\phi}}
\end{split}
\end{equation}

and

\begin{equation}\label{9}
\begin{split}
\{P_{\phi}, f(P_{a},P_{\phi})\}=-\theta\phi^{3}\frac{\partial f}{\partial P_{a}}
\end{split}
\end{equation}

By tending the parameter ($\theta$) to zero, all the above equations are recovered to standard commutative equations.
In the following, we apply the model mentioned in the previous section in the noncommutativity phase space presented in this section. We will provide an anisotropic constant-roll inflationary study using the mentioned Hamiltonian relationships and the field equations obtained from the above concepts.

\section{The Setup}

Here, the main aim is to achieve the Einstein-æther scalar-tensor cosmology field equations in the Noncommutative phase space.
Therefore, we first calculate the Hamilton according to equation (6). Then we will calculate the relationships and equations in the noncommutative space using canonical noncommutativity features into Poisson brackets. Finally, we calculate the field equations using the obtained equations.
From this relation, the Hamiltonian is calculated as follows.

\begin{equation}\label{10}
\begin{split}
\mathcal{H}=\frac{1}{N}\bigg(\frac{P_{\phi}^{2}}{2a^{3}}+\frac{846a^{5}(A-3B^{2})^{3}}{\big(aP_{a}-6BP_{\phi}\big)^{2}}\bigg)+Na^{3}V(\phi)
\end{split}
\end{equation}

In the above equations, parameters are seen, which will play an important role in the following calculations. ($P_{a}$) and ($P_{\phi}$) momenta conjugates of scale factor and momenta conjugates of the scalar field, respectively.
($\phi$) specifying the scalar field. Other parameters are also introduced in detail in the previous section.
To calculate the field equation in the noncommutative phase space, we use each of the parameters $\dot{a}$, $\dot{\phi}$, $\dot{P}_{a}$, and $\dot{P}_{\phi}$ with the Hamiltonian relation obtained in the above equation, i.e., in fact with a series of straightforward calculations and using equations (7- 10) and also using the property between the parameter and their
conjugates such as $\{a, P_{a}\}$ and $\{\phi, P_{\phi}\}$, we calculate the equations as follows.

\begin{equation}\label{11}
\begin{split}
\dot{a}=\{a, H\}=\frac{1}{N}\bigg[864a^{5}(A-3B^{2})^{3}\big(2a^{2}P_{a}-12aBP_{\phi}\big)^{-1}\bigg]
\end{split}
\end{equation}

\begin{equation}\label{12}
\begin{split}
\dot{\phi}=\frac{1}{N}\bigg[\frac{P_{\phi}}{a^{3}}+864a^{5}(A-3B^{2})^{3}\big(-12aBP_{a}+36B\big)^{-1}\bigg]
\end{split}
\end{equation}

\begin{equation}\label{13}
\begin{split}
&\dot{P}_{a}=\frac{1}{N}\bigg[\frac{3P_{\phi}^{2}}{2a^{4}}+\frac{\theta\phi^{3}}{a^{3}}P_{\phi}-\frac{864(A-3B^{2})^{3}5a^{4}}{(aP_{a}-6BP_{\phi})}\\
&+864a^{5}(A-3B^{2})^{3}\big(-2aP_{a}^{2}+12BP_{a}P_{\phi}-12aBP_{a}\theta\phi^{3}+36B\theta\phi^{3}\big)^{-1}\bigg]
\end{split}
\end{equation}

\begin{equation}\label{14}
\begin{split}
&\dot{P}_{\phi}=\frac{1}{N}\bigg[\frac{864a^{5}}{(aP_{a}-6BP_{\phi})^{2}}\big[-3A^{2}A'+18A^{2}BB'-108AB^{3}B'-27A'B^{4}+162B^{5}B'\big]\\
&+864a^{5}(A-3B^{2})^{3}\big(-2a^{2}\theta\phi^{3}P_{a}+12aB'P_{a}P_{\phi}+12aBP_{\phi}\theta\phi^{3}-36P_{\phi}B'\big)^{-1}\bigg]+Na^{3}V'
\end{split}
\end{equation}

\begin{equation*}\label{14}
\begin{split}
&\dot{P}_{N}=\{P_N, H\}=+\frac{P_{\phi}^{2}}{2a^{3}N^{2}}+\frac{846a^{5}(A-3B^{2})^{3}}{\big(aP_{a}-6BP_{\phi}\big)^{2}N^{2}}-a^{3}V(\phi)
\end{split}
\end{equation*}

Henceforth, we take the N = 1. Now according to the above equations and $\dot{P}_{N}=0$, we can calculate the Einstein-æther scalar-tensor cosmology  equations in the non-commutativity phase space in the following form.

\begin{equation}\label{15}
\begin{split}
3\big(\frac{\dot{a}}{a}\big)^{2}=8\pi G\big(\frac{1}{2}\dot{\phi}^{2}+V\big)
\end{split}
\end{equation}

\begin{equation}\label{16}
\begin{split}
\big(\frac{\ddot{a}}{a}\big)+2\big(\frac{\dot{a}}{a}\big)^{2}=-8\pi G\big(\frac{1}{2}\dot{\phi}^{2}+\frac{216(A-3B^{2})^{3}\theta\phi^{3}}{(A^{2}A'+B^{2}B')}\big)
\end{split}
\end{equation}

As it is clear from the above equations, we assume $8\pi G=1$; also, to calculate equation (15), we first squared both sides of equation (11). Then we put it in equation (14) by considering the equation (12), and after simplifying, the desired equation is calculated.
Also, to obtain equation (16), we first derivative equation (11) in terms of time, and then we use equations (12), (13), and (14).
Of course, as we have already mentioned, if the Non-commutativity parameter is assumed to be zero, the field equations become conventional equations in the literature.
As it is apparent in the field equations, parameters ($\theta$), A, and B play an important role in calculating cosmological parameters and quantities. In the following, we will calculate significant cosmological quantities such as Hubble parameter, potential, etc., using these field equations and the anisotropic constant-roll inflation scenario.

\section{Anisotropic constant-roll inflation, non-commutativity\\ \& Einstein-aether scalar-tensor cosmology}

In this section, we use equations (15) and (16); in this case, we will have.

\begin{equation}\label{17}
\begin{split}
\big(\frac{\ddot{a}}{a}\big)-\big(\frac{\dot{a}}{a}\big)^{2}=-\dot{\phi}^{2}-\frac{216(A-3B^{3})^{2}\theta\phi^{3}}{(A^{2}A'+B^{2}B')}
\end{split}
\end{equation}

Here we use the Hubble parameter definition $H=\frac{\dot{a}}{a}$ and rewrite equation (17) as follows.

\begin{equation}\label{18}
\begin{split}
\dot{H}=-\dot{\phi}^{2}-\frac{216(A-3B^{2})^{3}\theta\phi^{3}}{(A^{2}A'+B^{2}B')}
\end{split}
\end{equation}

Now using equations (18) and also $\dot{H}=\frac{dH}{d\phi}\dot{\phi}$, we will have.

\begin{equation}\label{19}
\begin{split}
\dot{\phi}^{2}+\frac{dH}{d\phi}\dot{\phi}+\frac{216(A-3B^{2})^{3}\theta\phi^{3}}{(A^{2}A'+B^{2}B')}
\end{split}
\end{equation}

We can calculate the parameter $\dot{\phi}$ as follows using a straightforward calculation.

\begin{equation}\label{20}
\begin{split}
\dot{\phi}=-\frac{1}{2}\frac{dH}{d\phi}\pm\frac{1}{2}\sqrt{\big(\frac{dH}{d\phi}\big)^{2}-\frac{864(A-3B^{2})^{3}\theta\phi^{3}}{(A^{2}A'+B^{2}B')}}
\end{split}
\end{equation}

We derivative from equation (20) in terms of time. So we will have.

\begin{equation}\label{21}
\begin{split}
&\ddot{\phi}=-\frac{1}{2}\frac{d^{2}H}{d\phi}^{2}\dot{\phi}\pm\bigg\{\frac{d^{2}H}{d\phi^{2}}\frac{dH}{d\phi}\dot{\phi}-\bigg[\bigg(864\big[3A^{2}A'\dot{\phi}-18AA'B^{2}\dot{\phi}-18BB'A^{2}\dot{\phi}+27B^{4}\dot{\phi}+108AB'B^{3}\dot{\phi}\\
&-162B'B^{5}\dot{\phi}\big]\theta\phi^{3}\times3\theta\phi^{2}\dot{\phi}(A-3B^{2})^{3}\bigg)\times(A^{2}A'\dot{\phi}+B^{2}B')-864(A-3B^{2})^{3}\theta\phi^{3}\bigg(A''A^{2}\dot{\phi}+2AA'^{2}\dot{\phi}\\
&+2BB'^{2}\dot{\phi}+B^{2}B''\dot{\phi}\bigg)\bigg]\bigg/\big((A^{2}A'+B^{2}B')\big)^{2}\bigg\}\bigg/
\bigg\{2\sqrt{\big(\frac{dH}{d\phi}\big)^{2}-\frac{864(A-3B^{2})^{3}\theta\phi^{3}}{(A^{2}A'+B^{2}B')}}\bigg\}
\end{split}
\end{equation}

With the integration of equation (21) and the well-known relation used in the constant-roll inflation scenario such as $\ddot{\phi}=-(3+\alpha)H\dot{\phi}$, we will have the following equation.

\begin{equation}\label{22}
\begin{split}
&-(3+\alpha)H=-\frac{1}{2}\frac{d^{2}H}{d\phi}^{2}\pm\bigg\{\frac{d^{2}H}{d\phi^{2}}\frac{dH}{d\phi}-\bigg[\bigg(864\big[3A^{2}A'-18AA'B^{2}-18BB'A^{2}+27B^{4}+108AB'B^{3}\\
&-162B'B^{5}\big]\theta\phi^{3}\times3\theta\phi^{2}(A-3B^{2})^{3}\bigg)\times(A^{2}A'+B^{2}B')-864(A-3B^{2})^{3}\theta\phi^{3}\bigg(A''A^{2}+2AA'^{2}\\
&+2BB'^{2}+B^{2}B''\bigg)\bigg]\bigg/\big((A^{2}A'+B^{2}B')\big)^{2}\bigg\}\bigg/\bigg\{2\sqrt{\big(\frac{dH}{d\phi}\big)^{2}-\frac{864(A-3B^{2})^{3}\theta\phi^{3}}{(A^{2}A'+B^{2}B')}}\bigg\}
\end{split}
\end{equation}

For the above equation, a suitable solution can be considered. That is if we consider parameters such as ($\theta$), A, and B as zero, equation (22) becomes an ordinary equation in the literature.
In fact, in this case, equation (22) becomes two equations, one zero and the other becomes the following form.

\begin{equation}\label{23}
\begin{split}
\frac{d^{2}H}{d\phi^{2}}-(3+\alpha)H=0
\end{split}
\end{equation}

The above equation is an ordinary differential equation whose answer will be calculated as follows.

\begin{equation}\label{24}
\begin{split}
H=c_{1}\exp\big(\sqrt{3+\alpha}\phi\big)+c_{2}\exp\big(-\sqrt{3+\alpha}\phi\big)
\end{split}
\end{equation}

Now we have to assume a particular ansatz to solve the whole equation (22) that contains essential parameters such as ($\theta$), A, and B so that the consequence of these parameters on significant quantities and parameters such as Hubble parameter and potential can be investigated. Hence we will have.

\begin{equation}\label{25}
\begin{split}
H=c_{1}\exp\big(\lambda(\theta, A, B)\sqrt{3+\alpha}\phi\big)+c_{2}\exp\big(-\lambda(\theta, A, B)\sqrt{3+\alpha}\phi\big)
\end{split}
\end{equation}

The parameter ($\lambda$) can be calculated directly by placing (25) in the equation (22).
By calculating this parameter ($\lambda$) and placing it in equation (25), we can calculate the explicit relationship for the Hubble parameter according to different boundary conditions. Then we can use it to calculate other quantities such as potential.

\subsection{Border condition I: $\big(c_{1}=c_{2}=\frac{M}{2}\big)$}
In this subsection, we apply the first boundary condition mentioned above to equation (25) with respect to equation (22).
In the following, different answers are obtained for the Hubble parameter, in which we consider only the positive solution.
Therefore, according to the concepts mentioned earlier, the Hubble parameter is calculated according to the first boundary condition in the following form.

\begin{equation}\label{26}
\begin{split}
&\mathcal{A}=3\sqrt{3}\bigg[\frac{2(3+\alpha)^{3/2}(-3+\sqrt{3}\phi+\sqrt{3+\alpha})^{2}(A''A^{2}+B^{2}B''+2AA'^{2}+2BB'^{2})}{27\sqrt{3}B^{4}B'+AA''+2AA'B^{2}+64AB''B^{3}}\\
&-\frac{\phi^{2}(3A^{2}A'+18AA'B^{2}\phi-162B'B^{5} \phi+9B'(-2+3B'+2B'\phi))\theta}{(A^{2}A'+B^{2}B')^{2}}\\
&+\frac{864(A-3B^{2})^{3}\theta \phi^{3}(2AA'^{2}+B(2B'^{2}+BB'))+A^{2}A'}{(A^{2}A'+B^{2}B')^{4}}\bigg]\\
&\mathcal{B}=2\sqrt{M^{2}\phi^{2}(3+\alpha)^{2}(-6-\alpha+2\sqrt{3}\sqrt{3+\alpha})}\\
&H=M\cosh\bigg[\frac{\phi\bigg(M(3+\alpha)-\frac{\mathcal{X}}{\mathcal{Y}}\bigg)}{2\sqrt{3}(3+\alpha)(M+M\sqrt{3+\alpha}/\sqrt{3})}\bigg]
\end{split}
\end{equation}

Now, using the above equation, we can calculate the potential for the first case, which is expressed in the following form.

\begin{equation}\label{27}
\begin{split}
&\mathcal{A}=3\sqrt{3}\bigg[\frac{2(3+\alpha)^{3/2}(-3+\sqrt{3}\phi+\sqrt{3+\alpha})^{2}(A''A^{2}+B^{2}B''+2AA'^{2}+2BB'^{2})}{27\sqrt{3}B^{4}B'+AA''+2AA'B^{2}+64AB''B^{3}}\\
&-\frac{\phi^{2}(3A^{2}A'+18AA'B^{2}\phi-162B'B^{5} \phi+9B'(-2+3B'+2B'\phi))\theta}{(A^{2}A'+B^{2}B')^{2}}\\
&+\frac{864(A-3B^{2})^{3}\theta \phi^{3}(2AA'^{2}+B(2B'^{2}+BB'))+A^{2}A'}{(A^{2}A'+B^{2}B')^{4}}\bigg]\\
&\mathcal{B}=2\sqrt{M^{2}\phi^{2}(3+\alpha)^{2}(-6-\alpha+2\sqrt{3}\sqrt{3+\alpha})}\\
&\mathcal{C}=3\sqrt{3}\big(\frac{4}{27}(3+\alpha)^{2}(-3+\sqrt{3}\phi\sqrt{3+\alpha}-\frac{\phi^{2}(2AA'-6B'+18BB')\theta}{(A^{2}A+B^{2}B')^{2}}\\
&-\frac{2\phi\theta\big(3A^{2}+2AA'\phi-6B'\phi+9B(-2+3B+2B'\phi)\big)}{(A^{2}A'+B^{2}B')^{2}}\\
&+\frac{3(A-2B^{2})^{3}\phi^{2}\theta\big(2A'^{2}A+B(2B'^{2}+BB'')+A^{2}A''\big)}{(A^{2}A'+B^{2}B')^{4}}\\
&V(\phi)=3M^{2}\cosh^{2}\bigg[\frac{\phi\bigg(M(3+\alpha)-\frac{\mathcal{A}}{\mathcal{B}}\bigg)}{2\sqrt{3}(3+\alpha)(M+M\sqrt{3+\alpha}/\sqrt{3})}\bigg]-2M^{2}\bigg(\frac{\phi}{2\sqrt{3}(3+\alpha)(M+M\sqrt{3+\alpha}/\sqrt{3})}\\
&\times\bigg(-\frac{\mathcal{C}}{\mathcal{B}}-\frac{3\sqrt{3}M^{2}\phi(3+\alpha)^{2}(-6-a+2\sqrt{3}\sqrt{3+\alpha})\mathcal{A}}{\mathcal{B}^{3}}\bigg)\\
&+\frac{M(3+\alpha)-\frac{\mathcal{A}}{\mathcal{B}}}{2\sqrt{3}(3+\alpha)(M+M\sqrt{3+\alpha}/\sqrt{3}}\bigg)^{2}\times\sinh^{2}\bigg[\frac{\phi\bigg(M(3+\alpha)-\frac{\mathcal{A}}{\mathcal{B}}\bigg)}{2\sqrt{3}(3+\alpha(M+M\sqrt{3+\alpha}/\sqrt{3}))}\bigg]
\end{split}
\end{equation}

Another important parameter, i.e., ($\dot{\phi}$), can also be calculated using the equations (25) and (20) as follows.

\begin{equation}\label{28}
\begin{split}
&\mathcal{A}=3\sqrt{3}\bigg[\frac{2(3+\alpha)^{3/2}(-3+\sqrt{3}\phi+\sqrt{3+\alpha})^{2}(A''A^{2}+B^{2}B''+2AA'^{2}+2BB'^{2})}{27\sqrt{3}B^{4}B'+AA''+2AA'B^{2}+64AB''B^{3}}\\
&-\frac{\phi^{2}(3A^{2}A'+18AA'B^{2}\phi-162B'B^{5} \phi+9B'(-2+3B'+2B'\phi))\theta}{(A^{2}A'+B^{2}B')^{2}}\\
&+\frac{864(A-3B^{2})^{3}\theta \phi^{3}(2AA'^{2}+B(2B'^{2}+BB'))+A^{2}A'}{(A^{2}A'+B^{2}B')^{4}}\bigg]\\
&\mathcal{B}=2\sqrt{M^{2}\phi^{2}(3+\alpha)^{2}(-6-\alpha+2\sqrt{3}\sqrt{3+\alpha})}\\
&\mathcal{C}=3\sqrt{3}\big(\frac{4}{27}(3+\alpha)^{2}(-3+\sqrt{3}\phi\sqrt{3+\alpha}-\frac{\phi^{2}(2AA'-6B'+18BB')\theta}{(A^{2}A+B^{2}B')^{2}}\\
&-\frac{2\phi\theta\big(3A^{2}+2AA'\phi-6B'\phi+9B(-2+3B+2B'\phi)\big)}{(A^{2}A'+B^{2}B')^{2}}\\
&+\frac{3(A-2B^{2})^{3}\phi^{2}\theta\big(2A'^{2}A+B(2B'^{2}+BB'')+A^{2}A''\big)}{(A^{2}A'+B^{2}B')^{4}}\\
&\dot{\phi}=-2M\bigg(\frac{\phi}{2\sqrt{3}(3+\alpha)M+M\sqrt{3+\alpha}/\sqrt{3}}\bigg[-\frac{\mathcal{C}}{\mathcal{B}}-\frac{3\sqrt{3}M^{2}\phi(3+\alpha)^{2}(-6-\alpha+2\sqrt{3}\sqrt{3+\alpha})\mathcal{A}}{\mathcal{B}^{3}}\bigg]\\
&\frac{M(3+\alpha)-\frac{\mathcal{A}}{\mathcal{B}}}{2\sqrt{3}(3+\alpha)(M+M\sqrt{3+\alpha}/\sqrt{3})}\bigg)-\bigg(-\frac{(A+3B^{2})^{3}}{(A^{2}A'+B^{2}B')}+4M^{2}\bigg(\frac{\phi}{2\sqrt{3}(3+\alpha)(M+M\sqrt{3+\alpha}/\sqrt{3})}\\
&\times\bigg[-\frac{\mathcal{C}}{\mathcal{B}}-\frac{3\sqrt{3}M^{2}\phi(3+\alpha)(-6-\alpha+2\sqrt{3}\sqrt{3+\alpha}/\sqrt{3})\mathcal{A}}{\mathcal{B}^{3}}\bigg]\\
&+\frac{M(3+\alpha)-\frac{\mathcal{A}}{\mathcal{B}}}{2\sqrt{3}(3+\alpha)(M+M\sqrt{3+\alpha}/\sqrt{3})}\bigg)^{2}\times\sinh^{2}\bigg[\frac{\phi\bigg(M(3+\alpha)-\frac{\mathcal{A}}{\mathcal{B}}\bigg)}{2\sqrt{3}(3+\alpha)(M+N\sqrt{3+\alpha}/\sqrt{3})}\bigg]\bigg)\bigg)^{1/2}
\end{split}
\end{equation}

In the following, other quantities such as scale factor can be calculated according to the definition of the Hubble parameter. In the continuation of this article, we apply the same calculations for the second boundary condition.

\subsection{Border condition II: $\big(c_{1}=\frac{M}{2}\big)$ and \big($c_{2}=-\frac{M}{2}\big)$}

In this subsection, we apply the second boundary condition to the equations (25) and (22), and the values for each of the parameters such as the Hubble parameter, potential, and $\dot{\phi}$ are calculated. Hence the Hubble parameter is calculated according to the second boundary condition as follows.

\begin{equation}\label{29}
\begin{split}
&\mathcal{X}=27\theta \phi\bigg((A^{2}A'+B^{2}B')\bigg(162A^{2}+18AA'\phi-6AB^{2}+9B'(-2AA'-12ABB'+3B'+36B'B^{5})\big)\bigg)\\
&-864(A+3B^{2})^{3}\phi+2B'\phi(2AA'^{2}+B(2B'^{2}+BB')+A^{2}A'')+\big(A''A^{2}+18AA'+B^{2}B''\phi)\\
&\times(-3AA'(-3+\alpha)^{2}\phi)\\
&\mathcal{Y}=M\sqrt{3+\alpha}\sqrt{M^{2}(3+\alpha)(-3+\phi^{2}(3+\alpha))}(A^{2}A'+B^{2}B')^{4}\\
&H=M\sinh\bigg[\frac{\phi\bigg((3+\alpha)+\frac{162M}{\sqrt{M^{2}(3+\alpha)(-3+\phi^{2}(3+\alpha)^{2})}}-\frac{\mathcal{X}}{\mathcal{Y}}\bigg)}{12\sqrt{3}(3+\alpha)}\bigg]
\end{split}
\end{equation}

According to the previous subsection, each of the parameters and quantities such as potential and $\dot{\phi}$ are calculated in the following form.

\begin{equation}\label{30}
\begin{split}
&\mathcal{X}=27\theta \phi\bigg((A^{2}A'+B^{2}B')\big(162A^{2}+18AA'\phi-6AB^{2}+9B'(-2AA'-12ABB'+3B'+36B'B^{5})\big)\bigg)\\
&-864(A+3B^{2})^{3}\phi+2B'\phi(2AA'^{2}+B(2B'^{2}+BB')+A^{2}A'')+\big(A''A^{2}+18AA'+B^{2}B''\phi)\\
&\times(-3AA'(-3+\alpha)^{2}\phi)\big)\\
&\mathcal{Y}=M\sqrt{3+\alpha}\sqrt{M^{2}(3+\alpha)(-3+\phi^{2}(3+\alpha))}(A^{2}A'+B^{2}B')^{4}\\
&\mathcal{Z}=27\theta\phi\bigg((2AA'-6B^{2}-12ABB'+36B'B^{4})(A^{2}A'+B^{2}B')^{2}\\
&-864(A-2B^{2})^{3}(2AA'^{2}+B(2B'^{2}+BB')+A^{2}A'')+2AA'^{2}+2BB'^{2}B''B^{2}+132AA'B^{2}\bigg)\bigg/\\
&\bigg(M\sqrt{3+\alpha}\sqrt{M^{2}(3+\alpha)(-3+\phi^{2}(3+\alpha)^{2})^{3/2}}(A^{2}A'+B^{2}B')^{4}\bigg)\\
&\mathcal{W}=\bigg(54M\phi^{2}(3+\alpha)^{3/2}(-3+\phi^{2}(3+\alpha))\theta((A^{2}A'+B^{2}B'))^{2}(3A^{2}+2AA'\phi-6B'\phi\\
&+9B(-2A'+3BB'+2B'\phi))-864(A-3B^{2})^{3}\phi(2AA'^{2}+B(2B'^{2}+BB')+A^{2}A'')\\
&+2AA'+162BB'^{2}+B^{2}B''\phi)\bigg)\bigg/\bigg(M^{2}(3+\alpha)(-3+\phi^{2}(3+\alpha)^{2})^{3/2}(A^{2}A'+B^{2}B')^{2}\bigg)\\
&V(\phi)=-2M^{2}\bigg(\frac{\phi}{12\sqrt{3}(3+\alpha)}\bigg[-\frac{6M^{3}\phi(3+\alpha)^{2}(-3+\phi^{2}(3+\alpha))}{(M^{2}(3+\alpha)(-3+\phi^{2}(3+\alpha))^{2})^{3/2}}-\mathcal{Z}+\mathcal{W}-\frac{\mathcal{X}}{\mathcal{Y}}\bigg]\\
&+\frac{12+(3+\alpha)+\frac{162M}{\sqrt{M^{2}(3+\alpha)(-3+\phi^{2}(3+\alpha)^{2})}}-\frac{\mathcal{X}}{\mathcal{Y}}}{12\sqrt{3}(3+\alpha)}\bigg)^{2}\times\cosh^{2}\bigg[\frac{\phi\bigg((3+\alpha)+\frac{162M}{\sqrt{M^{2}(3+\alpha)(-3+\phi^{2}(3+\alpha)^{2})}}-\frac{\mathcal{X}}{\mathcal{Y}}\bigg)}{12\sqrt{3}(3+\alpha)}\bigg]\\
&-3M^{2}\sinh^{2}\bigg[\frac{\phi\bigg((3+\alpha)+\frac{162M}{\sqrt{M^{2}(3+\alpha)(-3+\phi^{2}(3+\alpha)^{2})}}-\frac{\mathcal{X}}{\mathcal{Y}}\bigg)}{12\sqrt{3}(3+\alpha)}\bigg]
\end{split}
\end{equation}

and

\begin{equation}\label{31}
\begin{split}
&\mathcal{X}=27\theta \phi\bigg((A^{2}A'+B^{2}B')\big(162A^{2}+18AA'\phi-6AB^{2}+9B'(-2AA'-12ABB'+3B'+36B'B^{5})\big)\bigg)\\
&-864(A+3B^{2})^{3}\phi+2B'\phi(2AA'^{2}+B(2B'^{2}+BB')+A^{2}A'')+\big(A''A^{2}+18AA'+B^{2}B''\phi)\\
&\times(-3AA'(-3+\alpha)^{2}\phi)\big)\\
&\mathcal{Y}=M\sqrt{3+\alpha}\sqrt{M^{2}(3+\alpha)(-3+\phi^{2}(3+\alpha))}(A^{2}A'+B^{2}B')^{4}\\
&\mathcal{Z}=27\theta\phi\bigg((2AA'-6B^{2}-12ABB'+36B'B^{4})(A^{2}A'+B^{2}B')^{2}\\
&-864(A-2B^{2})^{3}(2AA'^{2}+B(2B'^{2}+BB')+A^{2}A'')+2AA'^{2}+2BB'^{2}B''B^{2}+132AA'B^{2}\bigg)\bigg/\\
&\bigg(M\sqrt{3+\alpha}\sqrt{M^{2}(3+\alpha)(-3+\phi^{2}(3+\alpha)^{2})^{3/2}}(A^{2}A'+B^{2}B')^{4}\bigg)\\
&\mathcal{W}=\bigg(54M\phi^{2}(3+\alpha)^{3/2}(-3+\phi^{2}(3+\alpha))\theta((A^{2}A'+B^{2}B'))^{2}(3A^{2}+2AA'\phi-6B'\phi\\
&+9B(-2A'+3BB'+2B'\phi))-864(A-3B^{2})^{3}\phi(2AA'^{2}+B(2B'^{2}+BB')+A^{2}A'')\\
&+2AA'+162BB'^{2}+B^{2}B''\phi)\bigg)\bigg/\bigg(M^{2}(3+\alpha)(-3+\phi^{2}(3+\alpha)^{2})^{3/2}(A^{2}A'+B^{2}B')^{2}\bigg)\\
&\dot{\phi}=-2M\bigg(\frac{\phi}{12\sqrt{3}(3+\alpha)}\bigg[-\mathcal{Z}-\frac{6M^{3}\phi(3+\alpha)^{2}(-3+\phi^{2}(3+\alpha))}{(M^{2}(3+\alpha)(-3+\phi^{2}(3+\alpha))^{2})^{3/2}}+\mathcal{W}-\frac{\mathcal{X}}{\mathcal{Y}}\bigg]\\
&+\frac{12+(3+\alpha)+\frac{162M}{\sqrt{M^{2}(3+\alpha)(-3+\phi^{2}(3+\alpha)^{2})}}-\frac{\mathcal{X}}{\mathcal{Y}}}{12\sqrt{3}(3+\alpha)}\bigg)\\
&\times\cosh^{2}\bigg[\frac{\phi\bigg((3+\alpha)+\frac{162M}{\sqrt{M^{2}(3+\alpha)(-3+\phi^{2}(3+\alpha)^{2})}}-\frac{\mathcal{X}}{\mathcal{Y}}\bigg)}{12\sqrt{3}(3+\alpha)}\bigg]-\bigg(\bigg[-\frac{(A-3B^{2})^{3}\theta\phi^{3}}{(A^{2}A'+B^{2}B')^{2}}\\
&+4M^{2}\bigg(\frac{\phi}{12\sqrt{3}(3+\alpha)}\bigg[-\frac{6M^{3}\phi(3+\alpha)^{2}(-3+\phi^{2}(3+\alpha))}{(M^{2}(3+\alpha)(-3+\phi^{2}(3+\alpha))^{2})^{3/2}}-\mathcal{Z}+\mathcal{W}-\frac{\mathcal{X}}{\mathcal{Y}}\bigg]\\
&+\frac{12+(3+\alpha)+\frac{162M}{\sqrt{M^{2}(3+\alpha)(-3+\phi^{2}(3+\alpha)^{2})}}-\frac{\mathcal{X}}{\mathcal{Y}}}{12\sqrt{3}(3+\alpha)}\bigg)^{2}\\
&\times\cosh^{2}\bigg[\frac{\phi\bigg((3+\alpha)+\frac{162M}{\sqrt{M^{2}(3+\alpha)(-3+\phi^{2}(3+\alpha)^{2})}}-\frac{\mathcal{X}}{\mathcal{Y}}\bigg)}{12\sqrt{3}(3+\alpha)}\bigg]\bigg]\bigg)^{1/2}
\end{split}
\end{equation}

\section{Concluding remarks}
In this paper, the primary purpose of this study was to investigate the constant-roll inflationary scenario with anisotropic conditions concerning the Einstein-aether Scalar-tensor Cosmology in noncommutative phase space.
That is, we first introduced an Einstein-aether scalar-tensor cosmological model. In this structure, in action integral of scalar-tensor,  one is introduced aether field with aether coefficients that it be a function of the scalar field, which is, in fact, a kind of extender of the previous Lorentz-violating theories.
Hence, we presented the point-like Lagrangian, which represents the equations of the  Einstein-aether Scalar-tensor model.
 Then we calculated the Hamiltonian of our model directly. According to the noncommutative phase space characteristics, we obtained the some equations of this model.
In the following, according to the constant-roll conditions, we studied the anisotropic constant-roll inflationary scenario and calculated some cosmological parameters of the mentioned model, such as the Hubble parameter, potential, etc. The findings of the mentioned paper can be extended to other scalar-tensor theories and challenge their cosmological applications.
The model mentioned in this paper or other models can also examine the scalar-tensor cosmology in different contexts and compare the results.
The findings of this article can be challenged with a new idea that has recently been very much of interest to researchers, namely to the swampland program, and compare the results with the latest observable data. It is also possible to study different theories of cosmology in the noncommutative phase space and select the best models among them that are most consistent with the latest observable data.
 Also, examining different types of cosmological models with such a proposed structure provided in this paper can propose a new classification for cosmological models.

\end{document}